# Phase space framework enables a variable-scale diffraction model for coherent imaging and display

ZHI LI,[1,2] XUHAO LUO,[3] JING WANG,[2] XIN YUAN,[4] DONGDONG TENG,[1,*] QIANG SONG,[2,**] AND HUIGAO DUAN[2, ***]

[1]*School of Physics, Sun Yat-sen University, Guangzhou 510220, China*
[2]*Greater Bay Area Institute for Innovation, Hunan University, Guangzhou 511300, China*
[3]*School of Physics Science and Engineering, Tongji University, Shanghai 200082, China*
[2]*School of Engineering, Westlake University, Hangzhou 330009, China*
*\*tengdd@mail.sysu.edu.cn*
*\*\*songqiangshanghai@foxmail.com*
*\*\*\*duanhg@hnu.edu.cn*

**Abstract:** The fast algorithms in Fourier optics have invigorated multifunctional device design and advanced imaging technologies. However, the necessity for fast computations has led to limitations in the widely used conventional Fourier methods, manifesting as fixed size image plane at a certain diffraction distance. These limitations pose challenges in intricate scaling transformations, 3D reconstructions and full-color displays. Currently, there is a lack of effective solutions, often resorting to pre-processing that compromise fidelity. In this paper, leveraging a higher-dimensional phase space method, we present a universal framework allowing for customized diffraction calculation methods. Within this framework, we establish a variable-scale diffraction computation model which allows the adjustment of the size of the image plane and can be operated by fast algorithms. We validate the model's robust variable-scale capabilities and its aberration automatic correction capability for full-color holography, achieving high fidelity. The large-magnification tomography experiment demonstrates that this model provides a superior solution for holographic 3D reconstruction. In addition, this model is applied to achieve full-color metasurface holography with near-zero crosstalk, showcasing its versatile applicability at nanoscale. Our model presents significant prospects for applications in the optics community, such as beam shaping, computer-generated holograms (CGHs), augmented reality (AR), metasurface optical elements (MOEs) and advanced holographic head-up display (HUD) systems.

## 1. Introduction

Traditional projection and display are no longer sufficient to meet the increasingly diverse demands, such as holographic head-up display (HUD) [1] and augmented reality (AR) [2]. Advanced imaging and display technologies, represented by Fourier optics with diffractive optics [3] as a prime example, provide an alternative approach. For instance, the advent of computer-generated hologram (CGH) has provided various manifestations for holography, including three-dimensional (3D) reconstruction [4–8], high-resolution capabilities [5,7,9,10], and implementations in augmented reality and virtual reality (VR) [7,11,12]. In recent years, driven by increasingly mature design and process capabilities, we have gained higher precision and flexibility in optimizing and fabricating devices. This has unlocked boundless possibilities in manufacturing, imaging and display, such as the utilization of freeform surface [13–15], diffractive optical element [13,16,17], optical waveguide [2,11] and liquid crystal [18,19]. In particular, the advent of metasurfaces has introduced a plethora of versatile modalities for coherent imaging [20–23]. Beyond the ultra-thin attributes inherent to traditional refractive devices, metasurfaces offer a substantial advantage in polarization control [24]. Metasurface holography, with its high information density in a planar format, holds tremendous potential for cutting-edge displays [23–28], paving the way for the realization of high-fidelity holography.

Despite continuous advancements in the design and fabrication methods of optical elements, without exception, these designs involving scalar diffractive optics invariably employ conventional Fourier methods, with angular spectrum method (ASM), single Fresnel transform (SFT), and their variants representing the prevailing approaches. Despite their widespread utilization and computational efficiency with fast Fourier transform (FFT), ASM and SFT suffer inherent limitations. Fixed sampling intervals, dictated by the conservation of the space-bandwidth product (SBP), restrict their universality. ASM requires equal-sized object and image planes, while SFT correlates image plane size linearly with propagation distance under the operation of Fourier optical transformation [3,29,30]. In a considerable proportion of cases, ASM based or SFT based diffraction iteration process may result in insufficient casting or resource wastage. Challenges arise when intricate scale transformation operations are required such as AR and holographic HUD, or when different depths in 3D projection demand varying levels of detail. Moreover, in color imaging and display, conventional Fourier methods such as Fraunhofer method (FM) may exhibit chromatic aberrations, necessitating complex correction and demanding additional computational resources. This leads to pixel pre-scaling for different colors during the computational process, compromising fidelity. In addition, the widely acknowledged accurate Rayleigh-Sommerfeld integral (RSI) is computationally slow and thus less suitable for inverse design applications. Conventional Fourier methods offers a rapid and efficient tool for diffraction calculation, with associated sampling and calculation strategies continually proposed and refined [4,31–37], but in general they merely represent a slight extension of conventional ASM and SFT, without breaking inherent scaling limitations. At times, new strategies accompanied by issues such as decreased image quality and narrowed the scope of application. For example, fast diffraction calculation algorithms can also be based on nonuniform FFT [38]. However, they may not be well-suited for inverse design applications, so they are generally rarely used. Overall, the modulation of light field still remains at a relatively low degree-of-freedom, even in cases of relatively uncomplicated coherent imaging and display, such as CGH imaging and automatic HUD. The fundamental reason for this lies in the fact that these methods are entrapped by the two local solutions of the Helmholtz equation in wave optics, and new solutions are extremely challenging to be delivered and difficult to achieve through fast algorithms.

To address these longstanding issues, we believe that continuing to find concise solutions directly from the Helmholtz equation may be a futile endeavor at present. Therefore, we seek breakthroughs from a higher-dimensional perspective to derive new diffraction computation methods suitable for fast algorithms. Phase space analysis, initially introduced in quantum mechanics through the behavior of Wigner distribution function (WDF) [39], has found extensive applications in describing optical systems [40,41]. It has been instrumental in understanding various phenomena, including the relationship between coherence and radiation measurement [42], the connection between ambiguity function and holography [43], and the measurement of partial coherence [44]. The completeness for signal description of WDF entirely exposes the spatial distribution of light completely in phase space. Albeit in higher dimensions, the WDF based phase space analysis itself constitutes a form of Fourier analysis. Phase space analysis directly analyzes the modulation of the light field from the perspective of distribution and transmission. Conventional Fourier optics can be viewed as a degenerate form of phase space analysis, and they are not mutually exclusive, as elaborated in the subsequent sections. In recent works [45–47], there have been some discussions on the relationship between conventional Fourier methods and WDF propagation. However, there have been few reports on fast algorithms designed for coherent light field diffraction calculations through phase space analysis.

In this paper, we chart a new course for coherent imaging computations within higher-dimensional phase space perspective. We first establish a universal framework that empowers the researchers to design diffraction calculation methods according to different application scenarios. Subsequently, we deliver a variable-scale model that allows the maximum spatial frequency of the image plane to be freely selected under certain constraints. This model still leverages Fourier analysis for fast computations, effectively addressing the size limitation issue inherent in conventional Fourier methods. The comparison and advantages of our model against other mainstream methods are illustrated in Table 1. Experimental validations, including advanced holographic and tomographic displays, demonstrate the model's robustness in variable-scale capability and chromatic aberration correction. Additionally, our model is applied to the design of full-color, near-zero crosstalk holographic metasurface, showcasing its applicability at nanoscale. Therefore, the longstanding algorithmic chromatic aberration issue in full-color holography would be effectively addressed. Rooted in phase space analysis, our approach offers a new tool for coherent imaging and display, providing an effective diffraction computation scheme for the diffractive optics community.

Table 1. Comparison of our model against the mainstream methods

| Model | Origin | Fast inverse algorithm | Working distance | Pre-processing | Scaling factor |
| --- | --- | --- | --- | --- | --- |
| RSI | Wave optics | No | Arbitrary | Not needed | 1 |
| ASM | Wave optics | Yes | Near field | Not needed | 1 |
| SFT | Wave optics | Yes | Far field | Required | Proportion of $\lambda z$ |
| FM | Wave optics | Yes | Ultra-far field | Required | Proportion of $\lambda z$ |
| Our Model | Phase space optics | Yes | Arbitrary | Not needed | Variable |

## 2. Universal framework and the proposed model

Under coherent illumination, a two-dimensional (2D) complex field distribution $U_o(r)$ possesses a WDF characterized in phase space. The WDF of $U_o(r)$ on the object or input plane is a four-dimensional (4D) function, encompassing information from both space and Fourier domains simultaneously, providing a comprehensive description of the light field signal:

$$W_o(\boldsymbol{r}_o, \boldsymbol{k}_o) = \iint_{-\infty}^{\infty} U_o(\boldsymbol{r}_o + \boldsymbol{r}_o'/2) U_o^*(\boldsymbol{r}_o - \boldsymbol{r}_o'/2) \exp(-i\boldsymbol{k}_o^t \cdot \boldsymbol{r}_o') \, d\boldsymbol{r}_o', \quad (1)$$

where $*$ denotes taking conjugation, $\boldsymbol{r}_o' = (\Delta x_o, \Delta y_o)^t$ is the spatial offset relative to $\boldsymbol{r}$ in space domain, and $\boldsymbol{k}_o = (k_{ox}, k_{oy})^t$ is the $k$-vector in Fourier domain with $t$ indicating the transpose operation. The point spread function (PSF) in free space characterizes the system's response to a point source, while the transfer function analyzed in the Fourier domain represents the response to plane waves. In phase space, since WDF is the complete signal representation, the propagation of it can respond to both point source and a single frequency, simultaneously and mathematically. It not only seamlessly combines the space domain and Fourier domain but also establishes a strong connection with the concept of "light rays" in geometrical optics. The corresponding ray spread function, which manifests as the response of the product of two Dirac delta functions $W_o(\boldsymbol{r}, \boldsymbol{k}) = \delta(\boldsymbol{r} - \boldsymbol{r}_o, \boldsymbol{k} - \boldsymbol{k}_o)$, forms a double Wigner distribution.

In scenarios where the Hamiltonian is limited to quadratic terms at most, the analysis of wave propagation can be elegantly achieved through straightforward matrix transformations. Linear canonical transform (LCT) emerges as a robust instrument for characterizing the behavior of light during its propagation within first-order optical systems [48]. By employing the paraxial approximation, the trajectories of light rays within such optical systems can be efficiently described using a fundamental ABCD matrix, which is a specific form of the LCTs. The ray transformation matrix $T_{4D} = [A, B; C, D]$ in phase space is applied as $[r_i; k_i] = T_{4D}[r_o; k_o] = [A, B; C, D][r_o; k_o]$, and the propagation relationship between $W_i$ on the image plane and $W_o$ on the object plane is expressed as [40,49]

$$W_i(Ar + Bk, Cr + Dk) = W_o(r, k). \tag{2}$$

In addition, the WDF transport equation of coherent light in free space is

$$(k/k)\frac{\partial W}{\partial r} + \left(\sqrt{(k^2 - |k|^2)}/k\right)\frac{\partial W}{\partial z} = 0, \tag{3}$$

where $z$ is the propagation distance, $\lambda$ is the wavelength and $k = 2\pi / \lambda$ denotes the wave number. Eq. (3) has the solution

$$W_i(r, k, z) = W_o\left(r - kz/\sqrt{(k^2 - |k|^2)}, k, 0\right). \tag{4}$$

Consider the low-band-limited characteristics of general objects and the applicability of sampling theorem, adapting $|k| \ll k$ and the transformation of $f = k / 2\pi$, $T_{4D}$ has the form

$$T_{4D} \cong \begin{bmatrix} I & 2\pi z k^{-1} I \\ 0 & I \end{bmatrix}, \tag{5}$$

where $I$ is the identity matrix. $T_{4D}$ contains the propagation characteristics of $W_o$ to $W_i$, having the ability to analyze the complete propagation pattern. For the sake of convenience in research, we adopt its 2D form, which can be effortlessly extended to four dimensions:

$$T_{2D} \cong \begin{bmatrix} 1 & \lambda z \\ 0 & 1 \end{bmatrix}. \tag{6}$$

$T_{2D}$ is a concise Fresnel transform matrix which is a two-order LCT. In the subsequent context, it will become apparent that this ostensibly straightforward matrix effectively encapsulates extensive Fourier analysis features [50]. Fourier transform and simple multiplication by quadratic-phase factors to $U_o(r)$ can be regarded as specific instances of the LCTs (detailed in Appendix A). Therefore, the LCTs are potent mathematical tools for analyzing diffraction phenomena. If we intend to conduct numerical computations for single diffraction or tasks such as phase retrieval in inverse design, it is essential to provide an appropriate matrix cascade while ensuring the conservation of the SBP to correspond to the respective computational operations. Meanwhile, the designed LCT matrix cascade which is employed for free-space diffraction presumably have the capability to degenerate into $T_{4D}$ under specific conditions. In essence, this leads to an incredible universal framework where we can extend an infinite number of methods to calculate light propagation, catering to diverse needs, especially in free space diffraction within first-order optical systems, and encompassing conventional Fourier optics. That is to say, in the phase space context, researchers or engineers can design diffraction computation methods independently and provide corresponding matrix cascade which must closely align with $T_{4D}$ or $T_{2D}$. Alternatively, $T_{4D}$ or $T_{2D}$ can be directly decomposed into the required computational operations, such as algorithms incorporating Fourier transforms.

To validate our perspective, an approach with higher-degree-of-freedom will be proposed to solve the calculation of coherent imaging and display, breaking through the size limitations of conventional Fourier methods represented by ASM and SFT. To facilitate the execution of FFT operation in either space domain and Fourier domain, $T_{2D}$ can be decomposed deliberately as follows:

$$T_{2D} = Q\left[\frac{m-1}{m\lambda z}\right] M[m] F^{-\pi/2}[1] Q\left[-\frac{\lambda z}{m}\right] F^{\pi/2}[1] Q\left[\frac{1-m}{\lambda z}\right], \tag{7}$$

where $Q[h] \cong \lim_{\epsilon \to 0}[1, \epsilon; h, 1 + \epsilon h] = [1,0; h, 1]$ and $h$ is a non-zero constant. $M[m] = [m, 0; 0, 1/m]$ is the injective scaling matrix with $m$ being the scaling factor, i.e., a magnifier.

$F^{\pi/2}[h] = [0, h; -1/h, 0]$ and $F^{-\pi/2}[h] = [0, -h; 1/h, 0]$ are the Fourier transform matrix and the inverse Fourier transform matrix, respectively. Phase space diagram (PSD) is defined as the region within phase space where WDF is non-negligible. In Fig. 1(a), a representative PSD is depicted, characterized by a spatial extent $L_o$ and a bandwidth $B_o$. Despite the potential for signal approximation through various methodologies [46], identifying explicit physical entities remains challenging. In this context, our focus is solely on illustrating the evolutionary trajectory of it. Fig. 1(a) shows the evolution process of the PSD, it can be clearly observed the process of clockwise (CW) rotation, counterclockwise (CCW) rotation, magnifying and shearing of the PSD under the application of each LCT according to Eq. (7). The same processes of ASM and SFT are easy to delivered (see Appendix A). The decomposition described above transforms a simple and abstract single phase space transformation into a series of chirp modulation and Fourier transformation operations, which is efficient and computationally tractable for implementation on a computer. Therefore, in space domain, the diffraction process in free space can be described as

$$U_i(x_i, m) = C Q_{p_3}(x_i, m) M\{[U_o(x_o) Q_{p_1}(x_o, m)] \otimes Q_{p2}(x_o, m)\}, \tag{8}$$

where $x_o$ and $x_i$ correspond to the coordinate systems at the object plane and the image plane, respectively. $C$ is a complex constant and $\otimes$ denotes linear convolution. The chirp modulator $Q_{p_1}(x_o, m) \cong \exp[i\varphi_{p1}(x_o, m)] = \exp[i\pi(1-m)x_o^2/\lambda z]$ is equivalent to $Q[(1-m)/\lambda z]$ in phase space, and similarly, $Q_{p_3}(x_i, m)$ is equivalent to $Q[(m-1)/m\lambda z]$. In addition, $Q_{p2}(x_o, m)$ corresponds to the space domain expression of $Q[-\lambda z/m]$. The injective transformation operator $M\{\cdot\}$ corresponds to $M[m]$ in phase space. Although $M\{\cdot\}$ does not result in any computational operations during numerical computation, in physical terms it directly maps the field distribution of $x_o$ system to $x_i$ system. This results in the calculated field distribution possessing the variable-scale characteristics. In the case of $m$ being equal to 1, Eq. (8) collapses into typical Fresnel form, and moreover, it shares similarities with Schmidt's description [32]. In the context of a fixed object plane with fixed sampling procedures, we retain the liberty to exert control over spatial extent of the image plane, i.e., the tunability of spatial frequency, as the SBP remains conserved. We refer to this method, allowing precise govern over spatial frequency, as the spatial frequency tunable method (SFTM). It is a tool with potential capabilities rooted in phase space analysis, applicable to solving problems in specific realms such as beam shaping, holography, metasurface design and solving pixel mismatch in end-to-end optimization [51]. It provides a powerful and easily implementable computational tool for the diffractive optics community, enabling them to break free from the confines of traditional methods. Fig. 1(b) demonstrates the variable-scale capability and the automatic aberration correction capability of SFTM.

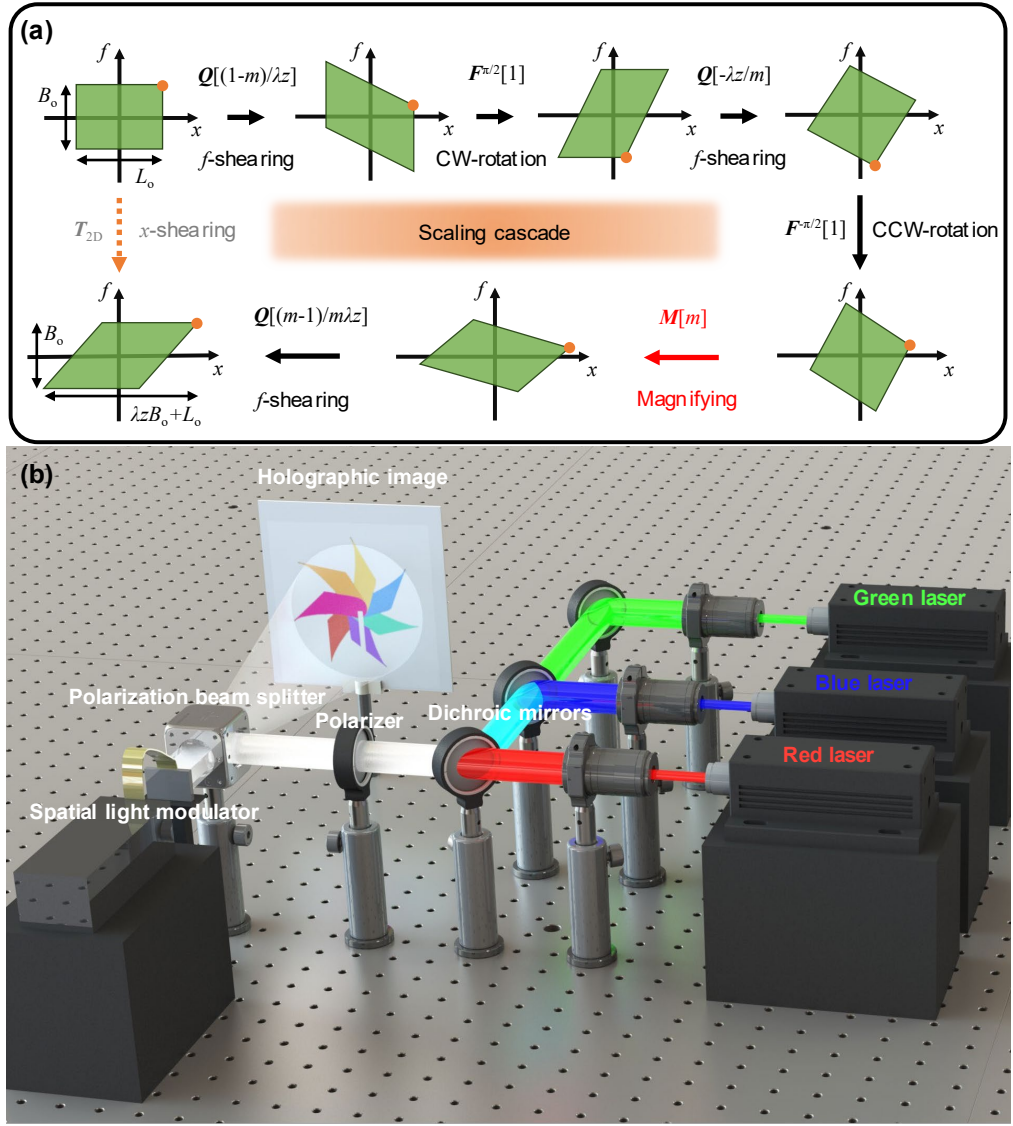

Fig.1. Modulation process of SFTM and demonstration of variable-scale holography. (a) Schematic diagram of evolution process of SFTM in phase space for $m > 1$. A typical PSD with a spatial extent $L_o$ and bandwidth $B_o$ is sheared in the $f$-direction through chirp modulation, and then undergoes coordinate transposition through simple Fourier transform. The PSD performs an inverse Fourier transform after shearing again and then magnifies by a magnifier. After the last chirp modulation, the PSD becomes the Fresnel form of its original state. (b) Demonstration of full-color holography without pre-processing using SFTM.

The relationships and constraints (CSTs) between $m$, $z$, and several other quantities are illustrated in Table 2 (the complete sampling criteria analysis is detailed in Appendix B). CST 4 offers two distinct sampling approaches, sampling in Fourier domain and sampling in space domain, to accommodate various scale transformation requirements. The partitioning of the sampling region, introduced by CST 4, is referred to as spatial frequency sampling region (SFSR) and space sampling region (SSR), respectively. Fig. 2 illustrates the permissible region of allowed $m$ and $z$ values for $\lambda = 0.532\ \mu m$, $N_0 = N/2 = 1000$, and $\delta x_o = 8\ \mu m$ scenarios. Meanwhile, the restrictions of both ASM and SFT have been delineated, where they represent only a segment of a straight line in the $m$-$z$ space. The threshold $z_0 = N_0(\delta x_o)^2 / \lambda$ is the boundary between ASM and SFT. It is evident that, with proper design, values for $m$ can be quite flexible, and it goes far beyond the conventional Fourier analysis represented by ASM, SFT, and their derivatives.

Table 2. Sampling CSTs for SFTM

| CST | Expression |
|---|---|
| CST 1 | $m \leq 1 + \lambda z / L_o \sqrt{(\delta x_o)^2 - (\lambda/2)^2}$ |
| CST 2 | $m \geq 1 - \lambda z / L_o \sqrt{(\delta x_o)^2 - (\lambda/2)^2}$ |
| CST 3 | $1 - \lambda z / (\delta x_o)^2 N_0 \leq m \leq 1 + \lambda z / (\delta x_o)^2 N_0$ |
| CST 4 | $\begin{cases} m \geq \lambda z / (\delta x_o)^2 N, \text{SFSR} \\ m \leq \lambda z / (\delta x_o)^2 N, \text{SSR} \end{cases}$ |
| CST 5 | $\begin{cases} 1/m \geq 1 - \lambda z / (\delta x_o)^2 N, \ m \geq 1 \\ m \geq 1 / 1 + \lambda z / (\delta x_o)^2 N, \ 0 < m \leq 1 \end{cases}$ |

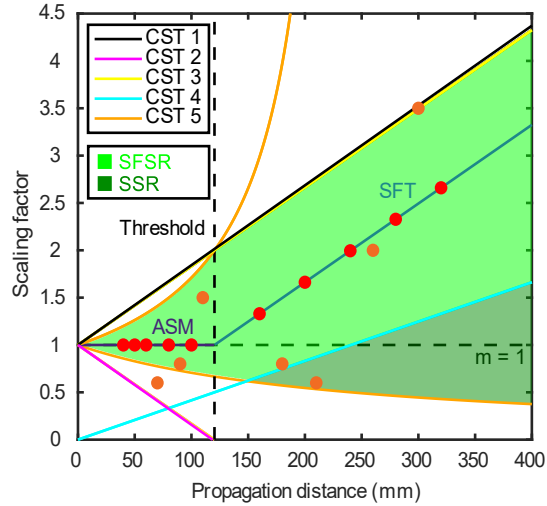

Fig. 2. Allowed $m$-$z$ space of SFTM for $\lambda = 0.532 \ \mu m$, $N_0 = N/2 = 1000$, and $\delta x_o = 8 \ \mu m$. The solid dots represent the data set used for the monochromatic CGH experiment.

## 3. Demonstration by CGHs

We employ a reflective 2D phase-only spatial light modulator (SLM) with a pixel size of $8 \ \mu m$ to verify the accuracy and practicality of SFTM. The inverse diffraction method has been employed to design 1000×1000-pixel CGHs. A slight perturbation deviation from the target will result in the "butterfly effect" in inverse diffraction after interactions [52]. Although the singularity of inverse diffraction kernels can be disregarded in the design process of CGHs used for display in the case of homogeneous light illumination, it is almost impossible to simulate and experiment with the correct image if the algorithm model is deviating from the exact $T_{4D}$. Traditional terminology refers to this situation as non-convergence. We utilize a three-stage iterative Fourier transform algorithm (IFTA) based on adaptive constraints in the Fourier domain (the algorithm flowchart is shown in Appendix C) instead of the traditional monotonous Gerchberg–Saxton (GS) algorithm to design CGHs with a signal window (SW) of 750×750 pixels. This choice is motivated partly by the susceptibility of the GS algorithm to local optima and partly due to practical considerations. The phase modulated by SLM is quantized into 256 steps, and achieving continuous control of phase is often impractical in the manufacturing process of diffractive optical elements. Importantly, the three-stage IFTA demonstrates superior reconstruction accuracy compared to the traditional GS algorithm, exhibiting higher values of the structural similarity index measure (SSIM) and lower values of root-mean-square error (RMSE) after hundreds of iterations.

Fig. 12 in Appendix F illustrates the experiment setup for the monochromatic display of the CGHs. Fig. 3(a) shows a comparison of the images projected by CGHs designed with ASM and SFTM within a propagation distance shorter than the threshold $z_0 = 120.3\ mm$. The scaling factor $m$ was set to 1 and it can be seen that within the oversampled region of transfer function for ASM at $z = 40\ mm$, $50\ mm$, $60\ mm$, $80\ mm$ and $100\ mm$ (specific parameter selections have been marked with solid points in Fig. 2), the rabbits designed with ASM and SFTM exhibits almost identical effects, without apparent aliasing or twin images. Similarly, as indicated by the solid points in the SFT regime in Fig. 2, we compared SFT and SFTM at the propagation distances of $z > z_0$, as Fig. 3(b) shows. Setting SFTM to have the same linear scaling characteristics as SFT, we captured the enlarged images at $z = 160\ mm$, $200\ mm$, $240\ mm$, $280\ mm$ and $320\ mm$, respectively. In Fig. 3(b), the remarkable congruence among the cat images, acquired through the three-stage IFTA with SFT and SFTM, is evident. This substantiates that the inverse algorithm, governed by SFTM, not only upholds $m = 1$ for $z < z_0$, where SFTM serves as a high-order alternative to ASM, but also showcases identical linear scaling characteristics to SFT for $z > z_0$. Consequently, SFTM can seamlessly substitute SFT in this regime. Additionally, holographic image projection experiments were conducted across different magnification ranges within the permissible regions of the $m$-$z$ space. As depicted in Fig. 3(c), a 0.8× chick and a 1.5× chick were captured at $z = 90\ mm$ and $z = 110\ mm$, respectively. Likewise, 0.8×, 0.6×, and 2.0× peacocks were successively projected to generate CGH images at $z = 180\ mm$, $210\ mm$, $260\ mm$.

In reality, most of the constraints of the sampling criteria originate from the Nyquist sampling theorem [3], which can be slightly relaxed in practical operations. On the other hand, $T_{4D}$ allows the size of the support on the image plane to extend to $L_o + \lambda z B_o$, not just the size of the image plane. By employing suitable filtering techniques, we can try to slightly break through the limitations of the allowed $m$-$z$ space for SFTM, leading to further break away from the limitations of conventional Fourier analysis. Similarly, in Fig. 3(c), a 0.6× chick out of the allowed $m$-$z$ space is projected at $z = 70\ mm$, and at another location of $z = 300\ mm$, a 3.5× peacock, just beyond the boundary of the constraints, is projected, and they both yield excellent imaging results. Furthermore, we assessed the SFTM algorithm's capability for long-distance projection by enlarging an image of a shark to 20 times its original size at $z = 1800\ mm$, as depicted in Fig. 3(d). In conventional Fresnel or Fraunhofer models, the achievable magnification factor is typically limited to around 15.

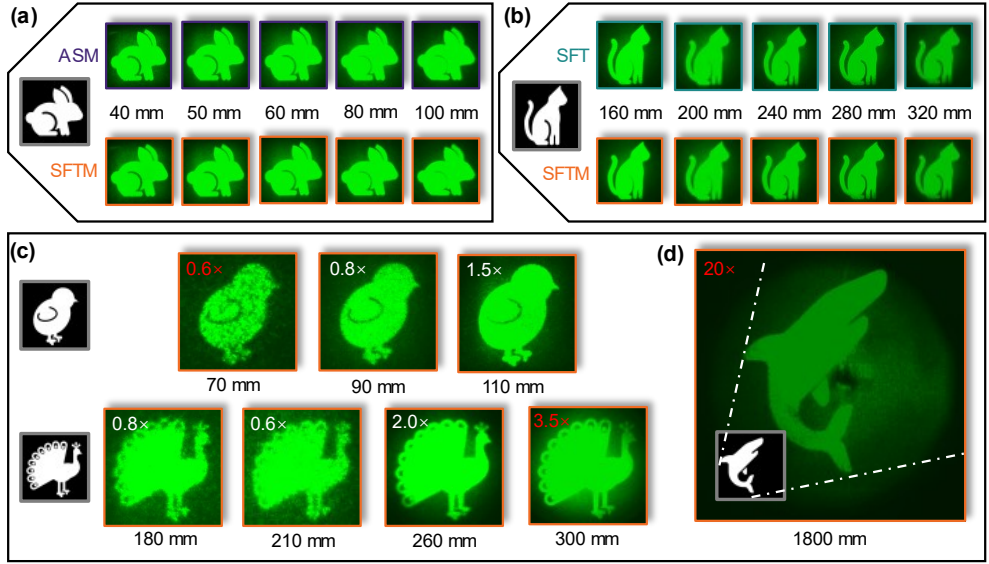

Fig. 3. Experimental results of SFTM algorithm under various circumstances. Comparison of SFTM to (a) ASM and (b) SFT illustrates that within a considerable diffraction distance, SFTM has almost the same effect within the applicable range of ASM and SFT. (c) The scaling ability of SFTM is presented under different $m$ and $z$. All $m$-$z$ values have been marked with orange solid dots in Fig. 1b. e Implementation of SFTM's long-distance and extreme magnification capability by projecting a 20× shark towards a distance of 1800 mm.

The far-field projection of holographic images conventionally relies on SFT based IFTA or the FM based IFTA. However, this approach introduces chromatic aberration, wherein the size of the image plane is directly proportional to the wavelength as depicted in Fig. 4(a). Addressing holographic image distortion within a large field of view (FOV) is approached through two methodologies: image pre-processing and hologram correction. Nevertheless, both of these methods are time-consuming and computationally intensive. Furthermore, they do not fundamentally alter the intrinsic linear-scale characteristics. As previously highlighted, our model is applicable for long-distance holographic image projection. By employing the SFTM based three-stage IFTA for holographic design, not only can we further minimize the crosstalk, but we can also rectify the linear-scale mismatch introduced by Fresnel-FFT algorithms. This approach eliminates the need for intricate and constrained pre-processing procedures. Moreover, the scale of the image plane is adjustable, offering substantial flexibility for the realization of full-color holographic displays. We initially decoupled a color windmill image based on the three primary colors, allowing them to iterate within their respective color channels. In Fig. 4(b), at a propagation distance of $z = 400\ mm$, we achieved a 3.5× magnification of a full-color holographic image projection. Notably, this process was conducted without involving chromatic aberration correction pre-processing. Remarkably, the full-color image exhibited a significant resemblance to the target.

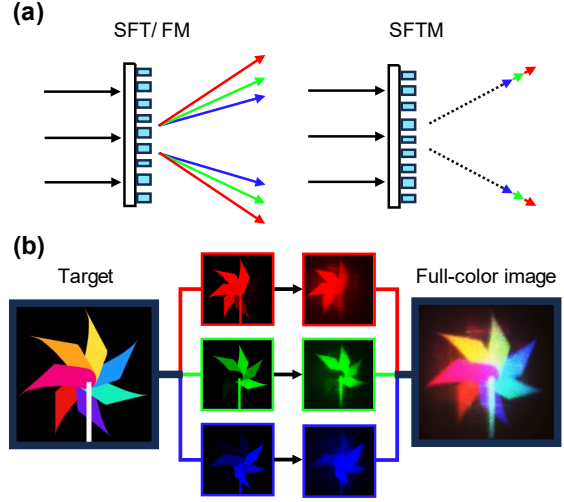

Fig. 4. Full-color holography without pre-processing. (a) Chromatic aberration comparison using SFT (Fraunhofer) algorithm and SFTM algorithm under vertical illumination in diffractive optical elements. (b) Implementation of full-color, automatic chromatic aberration correction in holographic images.

## 4. Implementation of tomography

During general 3D reconstruction process, different depths of a 3D object may require varying levels of detail. Typically, SFT is used for reconstructing large 3D objects, obtaining 3D Fresnel holograms, as the image plane size of SFT linearly increases with the propagation distance. However, since the size of an image plane at a particular depth remains fixed, while the 2D image at that depth may vary depending on the object and its position, pre-scaling of the 2D target images at each depth is generally required for 3D reconstruction. This inevitably leads to fidelity reduction in the total 3D scene. Recently, holograms of large objects with up to hundreds depth planes have been reported [4], offering an effective solution for future dynamic large-field 3D holography. However, this remarkable work is still based on the Fresnel method, requiring pre-processing of the 2D target image at each depth, which not only compromises fidelity but also consumes excessive computational resources. If our model can be utilized in tomography involving two or more image planes, then there will be significant potential for improvement in addressing these issues.

In the previous section, the scaling capabilities and chromatic aberration correction abilities of SFTM have been validated. Therefore, we conducted experiments to evaluate the multi-plane imaging capabilities of the SFTM algorithm, as shown in Fig. 5. Fig. 5(a) depicts a schematic illustration of the SFTM based tomographic reconstruction. In this representation, a fixed-size pixel on the hologram plane can be mapped to different image planes, each with pixels of varying sizes. This underscores the covariant capability of SFTM in tomographic reconstruction. Utilizing the tomography algorithm (see Appendix C for details), we imaged three pentagrams with different internal structures with magnification of 3×, 6× and 10× at depths of $z = 200\ mm$, $400\ mm$ and $700\ mm$, respectively. As shown in Fig. 5(b), a comparison of three identical duck objects reveals significant variations in magnification among the distinct pentagrams. For 3D Fresnel holograms, the depth of focus (DOF) of each image plane is a crucial factor influencing 3D reconstruction [4], and it is directly affected by $DOF_i \propto \lambda(z_i / N_{SW}\delta x_o)^2$ where $i$ is the number of the image plane. The requirement $z_{i+1} - z_i = \gamma(DOF_i + DOF_{i+1})$ for low crosstalk in image planes imposes a significant constraint on 3D reconstruction of 3D Fresnel holograms, where $\gamma$ is an empirical parameter. To mitigate this limitation, that is, to decrease DOFs, one relatively straightforward approach, without considering additional optimization or rearrangement of Fresnel zone plate phases, is to increase the resolution of the hologram. However, this method is often impractical and can lead to resource consumption and waste. We remind that, under the same projection depth, the increase in magnification leads to an elongation of DOF, as commonly known in the case of Fresnel holograms. Reflecting on Fig. 5(b), we observe that the 6× pentagram on the middle image plane is impacted by crosstalk from the 10× pentagram on the farthest image plane, whereas the 3× pentagram on the nearest image plane experiences minimal interference. The settings for these magnifications serve two purposes: firstly, to verify whether SFTM possesses sufficient diffraction modulation capability, and secondly, to facilitate the capture process. In numerous scenarios, achieving such extreme magnification may not be necessary. Therefore, it becomes viable to project images with varying spatial frequencies onto distant and multiple image planes. This approach can effectively surmount the constraints of 3D Fresnel holograms, opening avenues for the design of high-quality, super-multi-plane 3D variable-scale holograms.

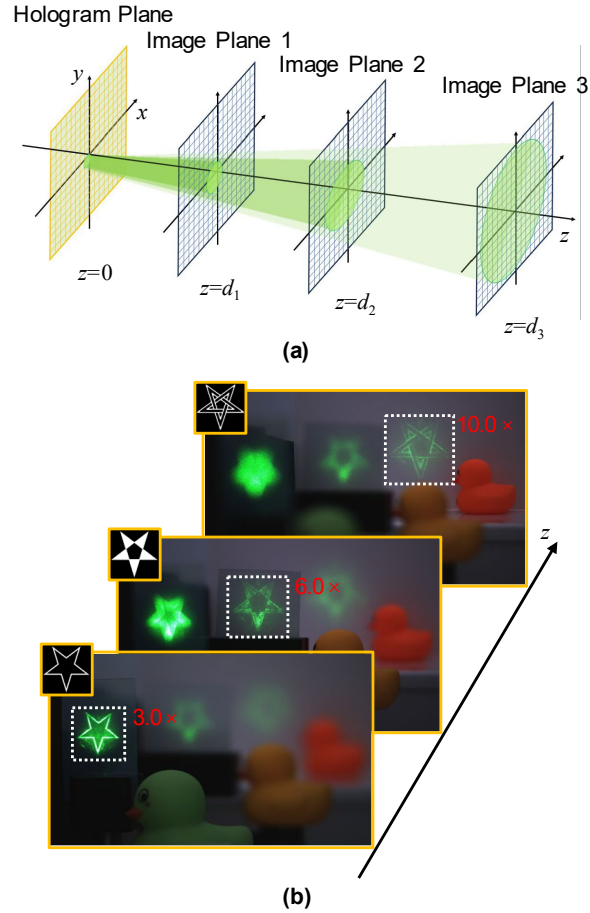

Fig. 5. Implementation of variable-scale tomography. (a) Schematic of SFTM based tomographic. The pixel sizes on three (or more) different image planes at different depths can be manipulated by SFTM algorithms as the SBP is conserved. (b) Three tomography images of 3×, 6×, and 10× pentagrams at depths of $z = 200\ mm$, $400\ mm$ and $700\ mm$ were projected, respectively.

## 5. Application in metasurface holography

The superiority of metasurfaces over traditional refractive and diffractive devices is evident not only in their ultra-thin features but also in their powerful capability for controlling multiple wavelengths and facilitating efficient polarization reuse. In our previous work [24], we implemented a TiO$_2$ metasurface design characterized by low-crosstalk and featured a tri-polarization-channel configuration for holographic display. Like most designs of full-color holographic metasurfaces, the previous design relies on the conventional Fraunhofer model, necessitating intricate pixel correction operations for different color channels. It will inevitably result in reduced resolution and fidelity. If the variable-scale model can be employed in context of metasurface holography to validate its applicability at nanoscale, this problem will be effectively resolved. Furthermore, researchers would have access to a more powerful tool for diffraction calculations when designing metasurfaces, freeing themselves from the constraints of ASM and SFT.

Fig. 6(a) illustrates the schematic of the near-zero crosstalk metasurface holography using the SFTM based variable-scale model. The letters "H", "N", and "U" corresponding to different colors should all have the same size. Fig. 6(b) illustrates the TiO$_2$ meta-atom design of metasurface, and the nanopillar which is capable of producing unique phases on mutually orthogonal linear polarizations is based on a square-shaped SiO$_2$ substrate with a period of $P = 400nm$, possessing three tunable degrees of freedom parameters $D_1$, $D_2$ and rotation angle $\theta$, with a fixed height of $H = 800nm$. The degree-of-freedom of the Jones matrix can be dynamically controlled by the three parameters of the meta-atom [24], achieving excellent broadband transmission and conversion efficiency characteristics of the metasurface. We decouple the light of the three colors, red ($\lambda = 0.633\ \mu m$), green ($\lambda = 0.532\ \mu m$), and blue ($\lambda = 0.450\ \mu m$), into different polarization channels, minimizing crosstalk between various polarization states. Then, at the image plane, they will superimpose based on the principles of the three primary colors. Leveraging the broadband transmission characteristics of the TiO$_2$ meta-atom, we set the polarization states of incident green and red light to *x*-polarized and blue light to *y*-polarized. The output red light is set to *x*-polarized, while green and blue lights are set to *y*-polarized. We decouple the target color image into red, green, and blue channels, allowing each channel component to undergo the three-stage IFTA process using the SFTM model for the reconstruction of the phase-only holograms (see Appendix D). The size of the metasurface is $400\mu m \times 400\mu m$, encompassing $N_0^2 = 1000 \times 1000$ units of meta-atoms. For the incident light of red, green, and blue, the allowed $m$-$z$ spaces are identified in Fig. S2. Significantly, when designing the metasurface, special consideration must be directed towards the allowed $m$-$z$ space for blue light incidence. This is particularly crucial since blue light exhibits the weakest diffraction capability, as constrained by the size of $L_o + \lambda_b z B_o$. Leveraging the lookup algorithm, we systematically identify and intricately arrange the appropriate meta-atoms with fixed parameters $D_1$, $D_2$ and $\theta$. This meticulous arrangement allows precise control over the Jones matrix, ensuring compliance with the specifications of three channels, diverse polarization states, and distinct phase distributions. As a result, this systematic procedure leads to the assembly of the metasurface.

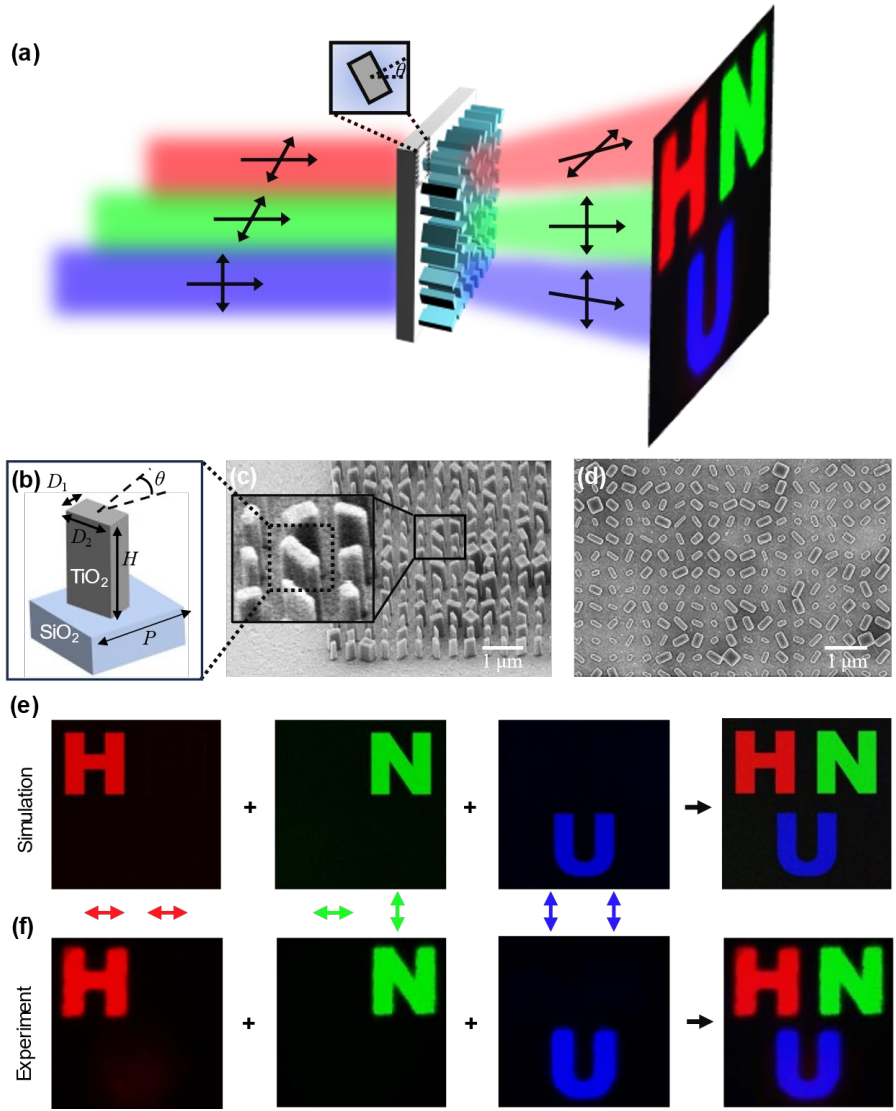

Fig. 6. Vectorial full-color SFTM based metasurface holography design. (a) Schematic of full-color metasurface holography. (b) A TiO$_2$ meta-atom with three independent tunable structure parameters of ($D_1$, $D_2$, $\theta$). Scanning electron microscopy (SEM) images of TiO$_2$ holographic metasurface in (c) oblique-view and (d) top-view are presented. (e) Simulation and (f) experiment results for "H", "N" and "U".

We employ the finite-difference time-domain (FDTD) method to craft a full-color holographic metasurface. Subsequently, the design is translated into sample through the utilization of electron beam lithography (EBL) followed by the reactive ion etching (RIE) process during the sample fabrication (see Appendix E for details). We designed a holographic metasurface with $m = 1.2$ and $z = 426.6 \mu m$, presenting independent three primary colors. At the image plane, the same-size letters "H", "N", and "U" were projected, corresponding to the red, green, and blue colors, respectively. Fig. 6(c)(d) shows the SEM images of the designed metasurface in oblique-view and top-view. As shown in Fig. 6(e)(f), the simulation and experimental results closely match, with minimal crosstalk between the color channels (The characterization setup for the metasurface sample can be found in Appendix F). The observable low crosstalk superposition capability among the three primary colors highlights the effectiveness of the meta-holography system. Our demonstration underscores the applicability of SFTM in nanoscale holography, effectively resolving the chromatic aberration issue caused by diffraction algorithms, thus paving the way for diverse applications in advanced coherent imaging and display.

## 6. Conclusion

The matrix cascade designed in our work provides powerful computational applications and extremely high-degree-of-freedom for diffraction calculations. It is evident that both the liquid crystal array in SLM and the meta-atom in the holography metasurface provide phase modulation capabilities spanning nearly 0 to $2\pi$. Consequently, the theoretical bandwidth of the WDF in phase space can be significantly broadened. Conventional IFTA is band-limited, explaining why there is not a "sufficiently large" region in the allowed $m$-$z$ space. Breaking this limitation is not infeasible, but doing so may result in a reduction in image quality. If we need to consider ultra-wideband scenarios, the WDFs of certain transfer functions or PSF may not be a simple Dirac delta function but could exhibit distorted tails on the space or frequency axis. In such cases, the highest order of the Hamiltonian might break the quadratic limitation, making the decomposition using LCTs exceedingly challenging. However, this is a separate topic, and we believe it holds significant importance in the shaping and optimization of the point spread function (or transfer function). Our variable-scale model and sampling criteria are robust enough to provide significant degree-of-freedom for most coherent imaging and displays, making it widely applicable in beam shaping, diffractive optical element design, CGH, 3D displays, HUDs and meta-holography.

In summary, from phase space perspective of WDF transportation characteristics, we have pioneered a novel trajectory in numerical computations for coherent imaging and display. This strategic approach allows us to intricately design matrix cascades by exploiting the nuanced interplay between LCTs and light field modulation. The resulting framework encompasses a spectrum of computational methodologies, including traditional Fourier methods. We derived a Fourier analysis method named SFTM with variable-scale diffraction regularity under the conservation of SBP, which offers much higher degrees of modulation freedom than traditional methods represented by ASM and SFT. We have demonstrated the effectiveness of the SFTM based algorithm, showcasing its powerful control of scaling capabilities and its automatic chromatic aberration correction capability, and point out that it may be a greater imaging implementation method than SFT based 3D Fresnel holography. Subsequently, we implemented a meta-atom design strategy establishing a direct correspondence with three distinct polarization channels and three primary color channels. By leveraging the three-stage IFTA based on SFTM, we achieved near-zero crosstalk, variable image scaling, and full-color metasurface holography, effectively resolving the longstanding issues of resolution and fidelity reduction in this field. This methodology offers a novel paradigm for coherent imaging and display. It is not overly complex compared to ASM and SFT, and it can be achieved by FFT-based algorithms, providing a powerful and efficient method for various diffraction calculations. Complemented by GPU acceleration, it exhibits the potential to realize dynamic, high degree-of-freedom and high frame-rate full-color displays in future applications.

## APPENDIX A: PHASE SPACE ANALYSIS

The WDF of $U(\boldsymbol{r})$ is real-valued but does not possess complete non-negativity, which prohibits it from being directly interpreted as an energy density. However, it can still be integrated along the spatial frequency vector, resulting in the intensity distribution for a two-dimensional optical field:

$$\iint_{-\infty}^{\infty} W(\boldsymbol{r}, \boldsymbol{f}) \, \mathrm{d}\boldsymbol{f} = U^*(\boldsymbol{r})U(\boldsymbol{r}). \tag{A1}$$

Alternatively, integrating along the spatial vector provides spectral information:

$$\iint_{-\infty}^{\infty} W(\mathbf{r}, \mathbf{f})\, d\mathbf{r} = \bar{U}^*(\mathbf{f})\bar{U}(\mathbf{f}). \tag{A2}$$

The symbol $*$ denotes taking conjugation, and $\bar{U}(\mathbf{f})$ is the Fourier transform of $U(\mathbf{r})$. Eq. (A1) and (A2) can serve as a direct representation of the diffraction pattern and be utilized for extracting spectral features.

For a two-component multiplicative signal $U_1(\mathbf{r}, \mathbf{f})U_2(\mathbf{r}, \mathbf{f})$, its WDF can be expressed as the convolution of the individual WDFs along the spatial frequency vectors for each component:

$$W_{1,2}(\mathbf{r}, \mathbf{f}) = W_1(\mathbf{r}, \mathbf{f}) \underset{f}{\otimes} W_2(\mathbf{r}, \mathbf{f}), \tag{A3}$$

where $\otimes$ represents the linear convolution. For spherical wave illumination or chirp modulated light field signals, their WDF in phase space can be simply regarded as the convolution of the original signal and Dirac delta function, which is convenient for characterizing the evolution process of PSD.

For a two-component convolved signal $U_1(\mathbf{r}, \mathbf{f}) \otimes U_2(\mathbf{r}, \mathbf{f})$, its WDF can be written as:

$$W_{1,2}(\mathbf{r}, \mathbf{f}) = W_1(\mathbf{r}, \mathbf{f}) \underset{r}{\otimes} W_2(\mathbf{r}, \mathbf{f}). \tag{A4}$$

The point spread function (PSF) of free space diffraction takes on a quadratic phase form. It can be drawn that the light field signal $U(\mathbf{r})$ on the object plane convolved with that PSF is equivalent to the convolution of WDF of $U(\mathbf{r})$ with the Dirac delta function along the $\mathbf{r}$-axis.

For the light field $U(\mathbf{r})$, the linear canonical transform has the following relationships with the transmission matrix $\mathbf{T} = [\mathbf{A}, \mathbf{B}; \mathbf{C}, \mathbf{D}]$ in phase space. Without considering the singularity of $\mathbf{B}$, in case that $\det \mathbf{B} \neq \mathbf{0}$,

$$U_{LCT}(\mathbf{r}_2) = (\det i\mathbf{B})^{1/2} \times$$
$$\iint_{-\infty}^{\infty} U(\mathbf{r}_1) \exp\left[i2\pi \left(\frac{1}{2}\mathbf{r}_2^t \mathbf{D}\mathbf{B}^{-1}\mathbf{r}_2 - \mathbf{r}_1 \mathbf{B}^{-1}\mathbf{r}_2 + \frac{1}{2}\mathbf{r}_1^t \mathbf{B}^{-1}\mathbf{A}\mathbf{r}_1\right)\right] d\mathbf{r}_1, \tag{A5}$$

else in the limiting case that $\mathbf{B} \to \mathbf{0}$,

$$U_{LCT}(\mathbf{r}_2) = |\det \mathbf{A}|^{-1/2} \exp[i\pi \mathbf{r}_2^t \mathbf{C}\mathbf{r}_1]U(\mathbf{r}_1). \tag{A6}$$

Since the multiplicative pure phase factors has no effect on the WDF of $U(\mathbf{r})$, if $\mathbf{B} = -\mathbf{C} = \mathbf{I}$, $\mathbf{A} = \mathbf{D} = \mathbf{0}$, it can be seen that Eq. (A5) is the Fourier transform of signal $U(\mathbf{r})$. Therefore, the "clockwise rotation" of matrix $\mathbf{T} = [\mathbf{0}, \mathbf{I}; -\mathbf{I}, \mathbf{0}]$ by $\pi / 2$ corresponds to the Fourier transform of $U(\mathbf{r})$, serving as a specific case of fractional Fourier transform. Similarly, "counterclockwise rotation" of matrix $\mathbf{T} = [\mathbf{0}, -\mathbf{I}; \mathbf{I}, \mathbf{0}]$ by $\pi / 2$ corresponds to the inverse Fourier transform operator. On the other hand, when considering the case of $\mathbf{B} \to \mathbf{0}$ and setting $\mathbf{A} = \mathbf{D} = \mathbf{I}$, according to Eq. (A6), chirp modulation will be obtained. The corresponding chirp modulation matrix is denoted as $\mathbf{T} = [\mathbf{I}, \mathbf{0}; h\mathbf{I}, \mathbf{I}]$, where $h$ is a non-zero constant. If we intend to magnify the WDF, or perform a scaling transformation, it is necessary for $\mathbf{C}$ to be zero according to Eq. (A6). As the conservation of SBP is guaranteed, $\mathbf{T} = [m\mathbf{I}, \mathbf{0}; \mathbf{0}, \mathbf{I}/m]$ represents a simple mapping with a scaling factor $m$, i.e., mapping the coordinate $\mathbf{r}_1$ to $\mathbf{r}_2 = m\mathbf{I}\mathbf{r}_1$. The mentioned LCT operators correspond to rotation, shearing, and stretch deformations for the PSD of $U(\mathbf{r})$ in phase space, respectively. In one-dimensional case, the situation becomes simpler and more intuitive.

ASM and SFT are the most widely used approaches for diffraction calculations, especially in cases involving inverse diffraction calculations. Considering the one-dimensional scenario, in phase space, ASM decomposes the transform matrix $\mathbf{T}_{2D} = [1, \lambda z; 0, 1]$ into

$$\boldsymbol{T}_{2D} = \begin{bmatrix} 0 & -1 \\ 1 & 0 \end{bmatrix} \begin{bmatrix} 1 & 0 \\ -\lambda z & 1 \end{bmatrix} \begin{bmatrix} 0 & 1 \\ -1 & 0 \end{bmatrix}, \qquad (A7)$$

and $\boldsymbol{T}_{2D}$ can also be decomposed by SFT as follows:

$$\boldsymbol{T}_{2D} = \begin{bmatrix} 1 & 0 \\ 1/\lambda z & 1 \end{bmatrix} \begin{bmatrix} 0 & \lambda z \\ -1/\lambda z & 0 \end{bmatrix} \begin{bmatrix} 1 & 0 \\ 1/\lambda z & 1 \end{bmatrix}. \qquad (A8)$$

Fig. 7. shows the evolution process of the PSD with a spatial extent $L_o$ and bandwidth $B_o$ through ASM and SFT. In addition, the PSF convolution method used for far-field calculations also has a similar process, the difference is that the PSD needs to be convolved with a "straight line" (at least under paraxial approximation) in phase space. These different methods ultimately lead to the same result, where PSD is horizontally sheared by $[1, \lambda z; 0,1]$.

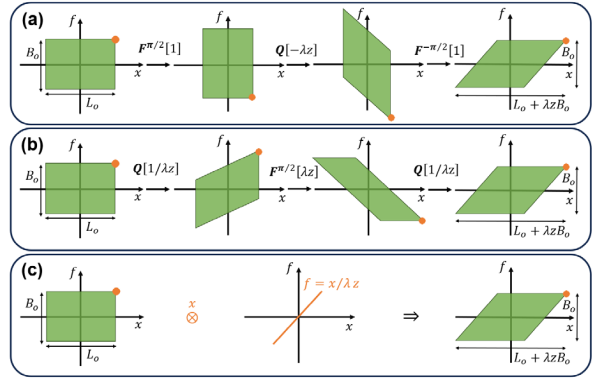

Fig. 7. Schematic diagram of PSD evolution process for (a)ASM, (b)SFT and (c)PSF convolution method in phase space.

## APPENDIX B: SAMPLING CRITERIA FOR SFTM

For the sake of convenience, the one-dimensional case will only be considered in the following discussion, and it can be readily extended to two dimensions. When the object support on the object plane is illuminated vertically by monochromatic plane waves, the maximum image support on the image plane is given by

$$L_i = L_o + 2z \tan \alpha_{max} = L_o + \frac{\lambda z}{\sqrt{(\delta x_o)^2 - (\lambda/2)^2}}, \qquad (B1)$$

where $L_o$ is the length of the object support, $L_i$ is the length of the maximum image support, $\lambda$ is the wave length of the plane wave and $z$ represents the propagation distance. Since the object support is band-limited, the maximum diffraction angle $\alpha_{max}$ is determined by $\alpha_{max} = \sin^{-1}(\lambda B_o / 2)$, where $B_o = 1 / \delta x_o$ is the bandwidth of the object.

Therefore, if the output field is calculated by SFTM, the support of the image plane should not exceed the maximum image support:

$$mL_o \leq L_o + \frac{\lambda z}{\sqrt{(\delta x_o)^2 - (\lambda/2)^2}}, \qquad (B2)$$

that is

$$m \leq 1 + \frac{\lambda z}{L_o \sqrt{(\delta x_o)^2 - (\lambda/2)^2}}. \qquad (B3)$$

Eq. (B3) is referred to as Constraint 1. Furthermore, it is worth mentioning that $\delta x_o > \lambda/2$ should be satisfied in inverse design, otherwise information will be lost in the form of evanescent waves.

When the image plane needs to be reduced in size, it cannot capture information from the entire extent of the object support if the object bandwidth $B_o$ is too narrow, that yields to

$$m \geq 1 - \frac{\lambda z}{L_o \sqrt{(\delta x_o)^2 - (\lambda/2)^2}} . \tag{B4}$$

Eq. (B4) is called Constraint 2.

In the case of SFTM being executed, conducting sampling in space domain or frequency domain primarily serves to mitigate the issue of inadequate sampling the quadratic phase factors. For the quadratic phase factor $Q_{p1} \propto \exp(i\varphi_{p1})$ which is directly multiplied by $U_o(x)$, the Nyquist sampling criterion requires

$$\delta x_o \left( \left| \frac{\partial \varphi_{p_1}}{\partial x_o} \right| \right)_{\max} \leq \pi, \tag{B5}$$

and it leads to

$$1 - \frac{\lambda z}{(\delta x_o)^2 N_0} \leq m \leq 1 + \frac{\lambda z}{(\delta x_o)^2 N_0} , \tag{B6}$$

where $N_0$ is the number of sampling points on the object support. Eq. (B6) is set to be Constraint 3.

For the sampling of the quadratic phase factor $Q_{p2}$, the criteria provide two distinct operational approaches to align with varying conditions. Because $Q_{p2}$ is described in Fourier domain, there are two fundamental methods for sampling it: directly sampling in Fourier domain, or sampling in space domain and then take the Fourier transform of it. The Nyquist sampling criterion provides completely opposite (or complementary) sampling requirements, and it can be derived that

$$\begin{cases} m \geq \frac{\lambda z}{(\delta x_o)^2 N}, & \text{sampling in Fourier domain} \\ m \leq \frac{\lambda z}{(\delta x_o)^2 N}, & \text{sampling in space domain} \end{cases}, \tag{B7}$$

where $N \geq N_0$ is the number of sampling points on the object plane. The two complementary inequalities in Eq. (B7) are called Constraint 4.

Up to this point, only the quadratic phase factor $Q_{p3}$ remains to be taken into consideration. If a single diffraction from free space is only required and solely concerned with the intensity pattern, then whether $Q_{p3}$ is well-sampled becomes irrelevant. However, if a certain work involves multiple diffraction and inverse diffraction calculations, with a focus on phase distributions, then the sampling of $Q_{p3}$ is critical and requires stringent attention. As per the analysis presented earlier in this text, applying the Nyquist sampling criterion, it leads to

$$\begin{cases} \frac{1}{m} \geq 1 - \frac{\lambda z}{(\delta x_o)^2 N}, & m \geq 1 \\ m \geq \frac{1}{1 + \lambda z / (\delta x_o)^2 N}, & 0 < m < 1 \end{cases}. \tag{B8}$$

Eq. (B8) is the proposed Constraint 5.

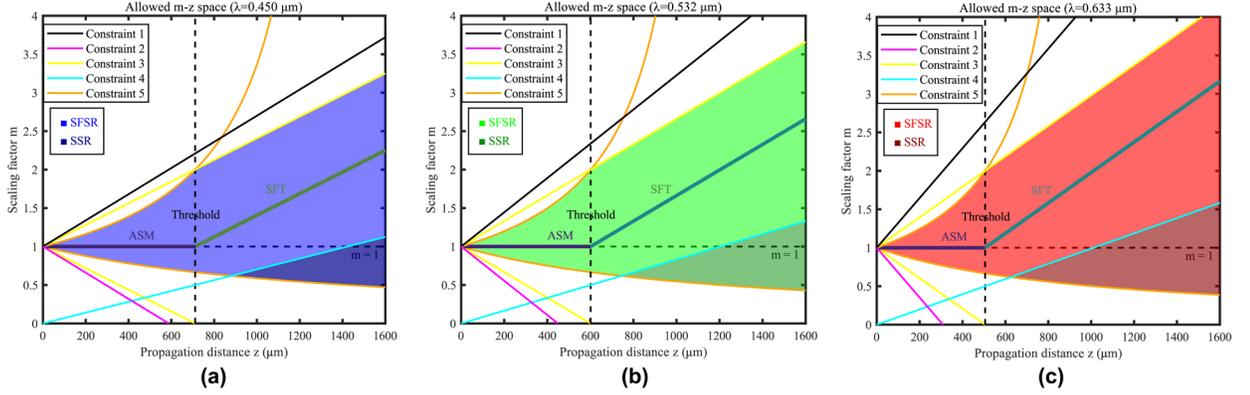

Fig. 8. The constraint relationship between $m$ and $z$ in cases (a) $\lambda = 0.450\ \mu m$, (b) $\lambda = 0.532\ \mu m$, and (c) $\lambda = 0.633\ \mu m$, respectively

Fig. 8(a)(b) and (c) respectively depict the allowed $m$-$z$ space at cases when $= 0.450\ \mu m$, $\lambda = 0.532\ \mu m$, and $\lambda = 0.633\ \mu m$, where $\delta x_o = 400\ nm$ and $N_0$ is set to 1000 as half of $N$. The cases intuitively demonstrate that the scaling range becomes more stringent as the wavelength of the incident light shortens. Simultaneously, it is readily apparent that the scaling range of SFTM is significantly broader than that of ASM and SFT.

When the LCTs are operated to the signal, the SBP in phase space remains constant. However, the support along both the space-axis and frequency-axis directions will expand or contract. Therefore, sampling points of $N = N_o$ on the object plane may potentially lead to insufficient sampling. In this article, uniform sampling is the primary method in SFTM because FFT can meet the vast majority of computational requirements. The number of sampling points should cover the SBP as comprehensively as possible. Consider a typical yet simple signal $\widehat{U}$, which is a simple rectangle PSD in phase space with bandwidth $B_o$ and spatial extent $L_o = N_o\ /\ B_o$, its four vertices in phase space are as follows

$$\boldsymbol{U} = \begin{bmatrix} x_1 & x_2 & x_3 & x_4 \\ f_1 & f_2 & f_3 & f_4 \end{bmatrix} = \begin{bmatrix} L_o\ /\ 2 & L_o\ /\ 2 & -L_o\ /\ 2 & -L_o\ /\ 2 \\ B_o\ /\ 2 & -B_o\ /\ 2 & -B_o\ /\ 2 & B_o\ /\ 2 \end{bmatrix}. \tag{B9}$$

Appling the LCTs to the signal $\widehat{U}$ in phase space, The Wigner distribution of $\widehat{U}$ in phase space will undergo affine transformation such as rotation and shearing. This directly leads to the number of sampling points $N$, which is under uniform sampling, being larger than $N_o$. Taking the four vertices of $\widehat{U}$ in phase space as an example, during the process of $\widehat{U}' = \widehat{L} \cdot \widehat{U}$, $\widehat{L}$ is a LCT operator with a LCT matrix $\boldsymbol{L} = [a, b; c, d]$, the bandwidth and the spatial extent change into

$$\boldsymbol{S}' = \begin{bmatrix} L' \\ B' \end{bmatrix} = \text{maxtaking}_{x,f}\{(\boldsymbol{LU})\boldsymbol{D}\}, \tag{B10}$$

where $\boldsymbol{D} = [1,1,1,0,0,0; -1,0,0,1,1,0; 0,-1,0,-1,0,1; 0,0,-1,0,-1,-1]$ is the distances matrix, and the operator $\text{maxtaking}_{x,f}\{\cdot\}$ is used to perform the operation of extracting the absolute maximum element from each row and placing its absolute value in the corresponding row of a new matrix [53]. For $\boldsymbol{U}$,

$$\boldsymbol{S}' = \begin{bmatrix} \max\{|aL_o|, |bB_o - aL_o|, |bB_o|, |bB_o + aL_o|\} \\ \max\{|cL_o|, |dB_o - cL_o|, |dB_o|, |dB_o + cL_o|\} \end{bmatrix}. \tag{B11}$$

Then, the sampling points on the object plane may not less than

$$N' = \frac{1}{2} \mathbf{S}' \begin{bmatrix} 0 & 1 \\ 1 & 0 \end{bmatrix} (\mathbf{S}')^t. \tag{B12}$$

Since SFTM does not involve fractional Fourier transforms, during the execution of SFTM, WDF or PSD in phase space will not undergo rotations that are non-integer multiples of $\pi/2$. In other words, rotations do not cause a change in the sampling points $N$. Additionally, the magnifier $[m, 0; 0, 1/m]$ will not impact on $N$. Judging from this, it can be seen that the quadratic phase factors is the primary determinant affecting $N$, so effectively addressing the shearing introduced by the quadratic phase factors during the SFTM operation is crucial for determining $N$. One of the advantages of analyzing the selection of $N$ in phase space is that, under the influence of LCTs, the deformation of the signal's WDF is intuitive, thereby avoiding intricate and stringent conditions in the space or Fourier domain.

Currently, $\mathbf{U}$ is employed once again to analyze the issue of selecting $N$. For $\mathbf{Q}_{p1} = \lim_{\epsilon \to 0}[1, \epsilon; (1-m)/\lambda z, 1 + \epsilon(1-m)/\lambda z] = \mathbf{Q}[(1-m)/\lambda z]$,

$$\mathbf{U}_1 = \mathbf{Q}_{p1}\mathbf{U} = \begin{bmatrix} L_o/2 & L_o/2 & -L_o/2 & -L_o/2 \\ \frac{B_o}{2} + \frac{(1-m)L_o}{2\lambda z} & -\frac{B_o}{2} + \frac{(1-m)L_o}{2\lambda z} & -\frac{B_o}{2} - \frac{(1-m)L_o}{2\lambda z} & \frac{B_o}{2} - \frac{(1-m)L_o}{2\lambda z} \end{bmatrix}. \tag{B13}$$

By applying Eq. (S2.10), (S2.11) and (S2.12), the required number of sampling points for the first chirp modulation satisfies

$$N_1 = L_o \left( \left| \frac{(1-m)L_o}{\lambda z} \right| + B_o \right). \tag{B14}$$

Similarly, for the remaining operations of the LCTs of SFTM, there are

$$N_2 = \frac{L_o + \lambda z B_o}{m} \left( \left| \frac{(1-m)L_o}{\lambda z} \right| + B_o \right), \tag{B15}$$

and

$$N_3 = (L_o + \lambda z B_o) B_o. \tag{S2.16}$$

Therefore, $N$ can be determined by

$$N \geq \max\{N_1, N_2, N_3\}. \tag{B17}$$

In general, once the bandwidth $B_o$ is fixed, $N$ can be determined by scaling factor $m$ and propagation distance $z$. Alternatively, we can adopt a "priori" approach by first determining $N$ and then assessing whether $m$ and $z$ meet the experimental requirements accordingly.

### APPENDIX C: ITERATIVE ALGORITHMS

As Fig. 9 shows, the three-stage IFTA consists of three stages [54]. The first stage involves the traditional GS algorithm, obtaining initial phases constrained by local optima. In the second stage, optimization of the SSIM of the diffraction pattern is achieved by minimizing a standard functional, i.e., the deviation between the actual diffraction image and the ideal image. This may involve a trade-off with diffraction efficiency, but for holographic image display, diffraction efficiency is not a critical metric as long as the diffraction effect is satisfactory. In the third stage, considering the practical manufacturing process of optical elements and the stepped nature of SLM's phase modulation, soft quantization is introduced to enhance the practicality of the algorithm. At the beginning, random phase can be attached to the object plane or to the image plane. In order to facilitate the implementation of the tomography, we uniformly attach the random phase to the object plane in this article, and the input for each stage after the first stage is the phase distribution of the object plane after the end of the previous stage.

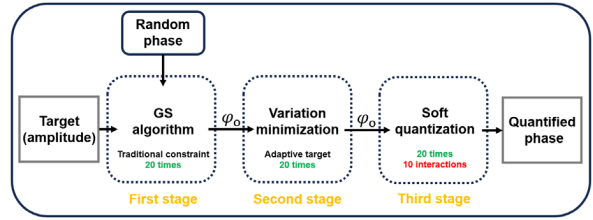

Fig. 9. Flowchart of the three-stage IFTA

Fig. 10 illustrates the algorithm flowchart for three-depth tomography. The workflow comprises three pivotal steps. The initial two steps align with the first two stages of the three-stage IFTA, while the concluding step executes a streamlined quantization process. Within each IFTA loop, holograms are generated for individual targets, followed by a weighted superposition achieved through multiplication by coefficient $\beta$, and $\varphi$ represents phase distribution. In each IFTA loop, initial complex amplitude holograms are generated using the three pentagram targets, and then perform complex superposition. The $\beta$ parameters are the complex amplitude superposition coefficients, which are empirical parameters used to improve the tomographic effect. To demonstrate the effectiveness of the variable-scale model, we actively reduced the degree-of-freedom of the algorithm by setting all $\beta$ parameters to 1/3.

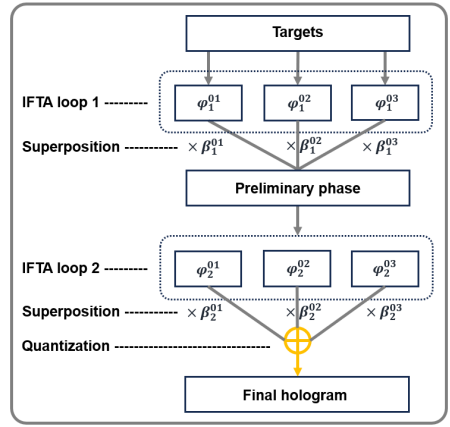

Fig. 10. Algorithm flowchart of three-plane tomography.

**APPENDIX D: DESIGN OF THE FULL-COLOR HOLOGRAPHIC METASURFACE**

We maintain the consistency of $m$ across the three channels to ensure uniform image plane sizes for all three channels. After obtaining the phase maps of each channel, we construct the metasurface by adaptively selecting meta-atoms that fulfill all corresponding polarization conditions and phases using a lookup algorithm. As the polarization states of incident and output light are determined, we present in Fig. 11(b) and (c) the relationships between the phase-change characteristics and transmission properties of green light with respect to the variations in $D_1$ and $D_2$. The parameters of the meta-atoms are highly flexible, as achieving phase variations from $-\pi$ to $\pi$ while maintaining high transmission properties is evidently achievable.

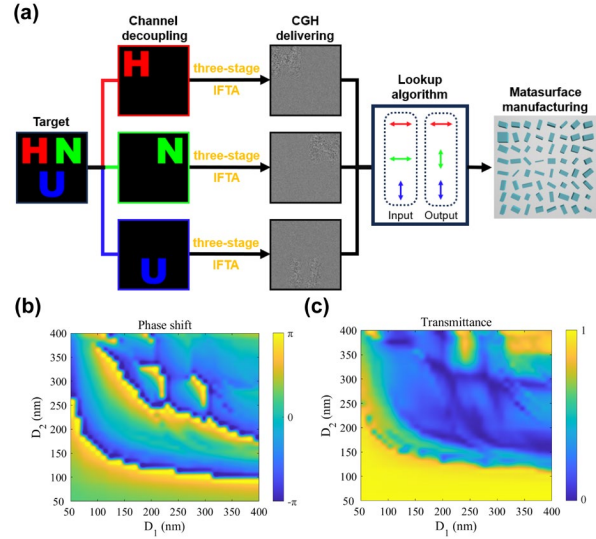

Fig. 11. (a) Algorithm flowchart of full-color metasurface holography. (b) phase shift and (c) Transmittance of green light obtained by sweep operations are shown.

**APPENDIX E: FABRICATION OF THE HOLOGRAPHIC METASURFACE**

First, an 800-nm-thick layer of polymethyl methacrylate (PMMA) electron beam resist was applied to the quartz substrate and baked on a hot plate at 180°C for 5 min. Subsequently, this resist layer was exposed by EBL (Raith 150$^{two}$) at a beam current of 200 pA. The sample was then immersed in a mixture of methyl isobutyl ketone (MIBK) and isopropyl alcohol (IPA) (MIBK: IPA = 1:3) for 1 min and left in IPA for 30 s. The reverse hole shape structure of the nanopillar was determined. Then, we deposited a 300-nm-thick $TiO_2$ layer using an atomic layer deposition system to fill the holes in the resist. In addition, there was a 300-nm-thick $TiO_2$ layer retained throughout the top of the sample. Therefore, we proceeded to etch the top $TiO_2$ layer using inductively coupled plasma reactive ion etching and remove the residual photoresist between the structures. Ultimately the tri-polarization-channel dielectric metasurface was obtained.

**APPENDIX F: EXPERIMENT SETUP**

This study utilized the HOLOEYE Photonics AG company's SLM with the model designation PLUTO-2.1 for conducting CGH experiments. We loaded phase-only holograms onto the SLM using MATLAB and software provided by HOLOEYE on a computer. Fig. 12 illustrates the experiment setup for the display of the CGHs. *S*-polarized collimated light with $\lambda = 0.532\ \mu m$, expanded, modulated and shaped, is incident onto the SLM. The reflected light modulated by SLM then passes through a 4*f* system and images on a CCD camera or a holographic screen after filtering.

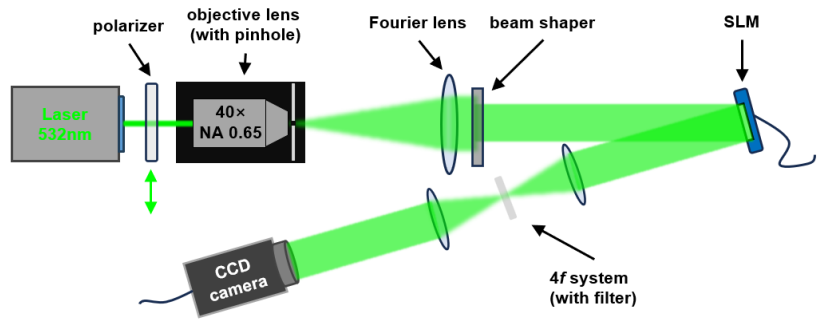

Fig. 12. Experiment setup with SLM

In the characterization of the holographic metasurface, we employed three lasers of red ($\lambda = 0.633\ \mu m$), green ($\lambda = 0.532\ \mu m$) and blue ($\lambda = 0.450\ \mu m$) for illumination. The three beams were combined using two dichroic mirrors and then vertically illuminated onto the metasurface, as illustrated in Fig. 13. The polarization states of the input and output of each laser beam are controlled by the polarizers and indicated by bidirectional arrows or dots. For convenient capturing, we utilized a 50× objective lens to magnify the holographic image. After a simple filtering process, a full-color holographic image was captured on a CCD camera.

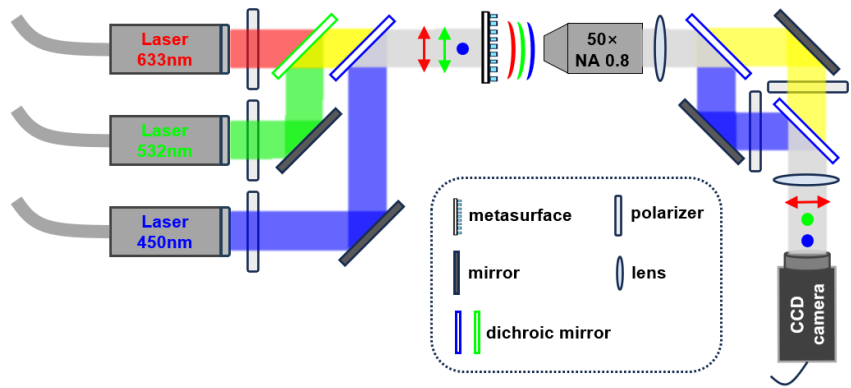

Fig. 13. Experiment setup for metasurface holography

**Funding.** Guangzhou Major R&D Funds (ZF202200125); Guangdong Province Key R&D Projects (2019B010154002)

**Disclosures.** The authors declare no conflicts of interest.

**Data availability.** Data can be obtained from the corresponding author upon reasonable request.

## References

1. J. Skirnewskaja and T. D. Wilkinson, "Automotive Holographic Head-Up Displays," Adv. Mater. **34**, 2110463 (2022).
2. Y. Ding, Q. Yang, Y. Li, Z. Yang, Z. Wang, H. Liang, and S.-T. Wu, "Waveguide-based augmented reality displays: perspectives and challenges," eLight **3**, 24 (2023).

3. J. W. Goodman, *Introduction to Fourier Optics* (Roberts and Company publishers, 2005).
4. G. Makey, Ö. Yavuz, D. K. Kesim, A. Turnalı, P. Elahi, S. Ilday, O. Tokel, and F. Ö. Ilday, "Breaking crosstalk limits to dynamic holography using orthogonality of high-dimensional random vectors," Nat. Photonics **13**, 251–256 (2019).
5. D. Yang, W. Seo, H. Yu, S. I. Kim, B. Shin, C.-K. Lee, S. Moon, J. An, J.-Y. Hong, G. Sung, and H.-S. Lee, "Diffraction-engineered holography: Beyond the depth representation limit of holographic displays," Nat. Commun. **13**, 6012 (2022).
6. L. Shi, B. Li, and W. Matusik, "End-to-end learning of 3D phase-only holograms for holographic display," Light Sci. Appl. **11**, 247 (2022).
7. D. Pi, J. Liu, and Y. Wang, "Review of computer-generated hologram algorithms for color dynamic holographic three-dimensional display," Light Sci. Appl. **11**, 231 (2022).
8. A. H. Dorrah, P. Bordoloi, V. S. de Angelis, J. O. de Sarro, L. A. Ambrosio, M. Zamboni-Rached, and F. Capasso, "Light sheets for continuous-depth holography and three-dimensional volumetric displays," Nat. Photonics 1–8 (2023).
9. S. Schaffert, B. Pfau, J. Geilhufe, C. M. Guenther, M. Schneider, C. von K. Schmising, and S. Eisebitt, "High-resolution magnetic-domain imaging by Fourier transform holography at 21 nm wavelength," New J. Phys. **15**, 093042 (2013).
10. L. Shi, B. Li, C. Kim, P. Kellnhofer, and W. Matusik, "Towards real-time photorealistic 3D holography with deep neural networks," Nature **591**, 234–239 (2021).
11. Z. Huang, D. L. Marks, and D. R. Smith, "Out-of-plane computer-generated multicolor waveguide holography," Optica **6**, 119–124 (2019).
12. A. Maimone, A. Georgiou, and J. S. Kollin, "Holographic near-eye displays for virtual and augmented reality," ACM Trans. Graph. **36**, 1–16 (2017).
13. O. Mendoza-yero, "Dynamic freeform diffractive lens," Optica **10**, 443–449 (2023).
14. F. Duerr and H. Thienpont, "Freeform imaging systems: Fermat's principle unlocks "first time right" design," Light Adv. Manuf. **10**, 1029–1040 (2021).
15. S. Kumar, Z. Tong, and X. Jiang, "Advances in the design and manufacturing of novel freeform optics," Int. J. Extreme Manuf. **4**, 032004 (2022).
16. Y. Lee, M. J. Low, D. Yang, H. K. Nam, T.-S. D. Le, S. E. Lee, H. Han, S. Kim, Q. H. Vu, H. Yoo, H. Yoon, J. Lee, S. Sandeep, K. Lee, S.-W. Kim, and Y.-J. Kim, "Ultra-thin light-weight laser-induced-graphene (LIG) diffractive optics," Light Sci. Appl. **12**, 146 (2023).
17. S. Schmidt, S. Thiele, A. Toulouse, C. Boesel, T. Tiess, A. Herkommer, H. Gross, and H. Giessen, "Tailored micro-optical freeform holograms for integrated complex beam shaping," Optica **7**, 1279–1286 (2020).
18. K. Yin, E.-L. Hsiang, J. Zou, Y. Li, Z. Yang, Q. Yang, P.-C. Lai, C.-L. Lin, and S.-T. Wu, "Advanced liquid crystal devices for augmented reality and virtual reality displays: principles and applications," Light Sci. Appl. **11**, 161 (2022).

19. Z. Luo, Y. Li, J. Semmen, Y. Rao, and S.-T. Wu, "Achromatic diffractive liquid-crystal optics for virtual reality displays," Light Sci. Appl. **12**, 230 (2023).
20. Z. Liu, D. Wang, H. Gao, M. Li, H. Zhou, and C. Zhang, "Metasurface-enabled augmented reality display: a review," Adv. Photonics **5**, 034001–034001 (2023).
21. X. Li, Q. Chen, X. Zhang, R. Zhao, S. Xiao, Y. Wang, and L. Huang, "Time-sequential color code division multiplexing holographic display with metasurface," Opto-Electron. Adv. **6**, 220060 (2023).
22. A. H. Dorrah and F. Capasso, "Tunable structured light with flat optics," Science **376**, 367-+ (2022).
23. J. Kim, J. Seong, Y. Yang, S.-W. Moon, T. Badloe, and J. Rho, "Tunable metasurfaces towards versatile metalenses and metaholograms: a review," Adv. Photonics **4**, 024001–024001 (2022).
24. Y. Hu, L. Li, Y. Wang, M. Meng, L. Jin, X. Luo, Y. Chen, X. Li, S. Xiao, H. Wang, Y. Luo, C.-W. Qiu, and H. Duan, "Trichromatic and Tripolarization-Channel Holography with Noninterleaved Dielectric Metasurface," Nano Lett. **20**, 994–1002 (2020).
25. J. Yang, S. Gurung, S. Bej, P. Ni, and H. W. Howard Lee, "Active optical metasurfaces: comprehensive review on physics, mechanisms, and prospective applications," Rep. Prog. Phys. **85**, 036101 (2022).
26. H. Yang, P. He, K. Ou, Y. Hu, Y. Jiang, X. Ou, H. Jia, Z. Xie, X. Yuan, and H. Duan, "Angular momentum holography via a minimalist metasurface for optical nested encryption," Light Sci. Appl. **12**, 79 (2023).
27. S. So, J. Kim, T. Badloe, C. Lee, Y. Yang, H. Kang, and J. Rho, "Multicolor and 3D Holography Generated by Inverse-Designed Single-Cell Metasurfaces," Adv. Mater. **35**, (2023).
28. Q. Song, X. Liu, C.-W. Qiu, and P. Genevet, "Vectorial metasurface holography," Appl. Phys. Rev. **9**, (2022).
29. D. Mendlovic, Z. Zalevsky, and N. Konforti, "Computation considerations and fast algorithms for calculating the diffraction integral," J. Mod. Opt. **44**, 407–414 (1997).
30. D. G. Voelz and M. C. Roggemann, "Digital simulation of scalar optical diffraction: revisiting chirp function sampling criteria and consequences," Appl. Opt. **48**, 6132–6142 (2009).
31. K. Matsushima and T. Shimobaba, "Band-limited angular spectrum method for numerical simulation of free-space propagation in far and near fields," Opt. Express **17**, 19662–19673 (2009).
32. J. D. Schmidt, *Numerical Simulation of Optical Wave Propagation with Examples in MATLAB* (SPIE, 2010).
33. W. Zhang, H. Zhang, and G. Jin, "Band-extended angular spectrum method for accurate diffraction calculation in a wide propagation range," Opt. Lett. **45**, 1543–1546 (2020).
34. Y. Hu, Z. Wang, X. Wang, S. Ji, C. Zhang, J. Li, W. Zhu, D. Wu, and J. Chu, "Efficient full-path optical calculation of scalar and vector diffraction using the Bluestein method," Light Sci. Appl. **9**, 119 (2020).

35. W. Zhang, H. Zhang, and G. Jin, "Frequency sampling strategy for numerical diffraction calculations," Opt. Express **28**, 39916–39932 (2020).
36. H. Wei, X. Liu, X. Hao, E. Y. Lam, and Y. Peng, "Modeling off-axis diffraction with the least-sampling angular spectrum method," Optica **10**, 959–962 (2023).
37. H. Yu, Y. Kim, D. Yang, W. Seo, Y. Kim, J.-Y. Hong, H. Song, G. Sung, Y. Sung, and S.-W. Min, "Deep learning-based incoherent holographic camera enabling acquisition of real-world holograms for holographic streaming system," Nat. Commun. **14**, 3534 (2023).
38. J.-Y. Lee and L. Greengard, "The type 3 nonuniform FFT and its applications," Journal of Computational Physics **206**, 1–5 (2005).
39. E. Wigner, "On the Quantum Correction For Thermodynamic Equilibrium," Phys. Rev. **40**, 749–759 (1932).
40. M. Testorf, B. Hennelly, and J. Ojeda-Castañeda, *Phase-Space Optics: Fundamentals and Applications* (McGraw-Hill Education, 2010).
41. M. A. Alonso, "Wigner functions in optics: describing beams as ray bundles and pulses as particle ensembles," Adv. Opt. Photonics **3**, 272–365 (2011).
42. A. Walther, "Propagation of the generalized radiance through lenses," J. Opt. Soc. Am. **68**, 1606–1610 (1978).
43. G. Situ and J. T. Sheridan, "Holography: an interpretation from the phase-space point of view," Opt. Lett. **32**, 3492–3494 (2007).
44. L. Waller, G. Situ, and J. W. Fleischer, "Phase-space measurement and coherence synthesis of optical beams," Nat. Photonics **6**, 474–479 (2012).
45. A. Stern and B. Javidi, "Sampling in the light of Wigner distribution," J. Opt. Soc. Am. A **21**, 360 (2004).
46. J. Xiao, W. Zhang, and H. Zhang, "Sampling analysis for Fresnel diffraction fields based on phase space representation," J. Opt. Soc. Am. A **39**, A15–A28 (2022).
47. J. Xiao, W. Zhang, and H. Zhang, "Inverse diffraction in phase space," J. Opt. Soc. Am. A **40**, 175–184 (2023).
48. J. J. Healy, M. Alper Kutay, H. M. Ozaktas, and J. T. Sheridan, eds., *Linear Canonical Transforms: Theory and Applications*, Springer Series in Optical Sciences (Springer New York, 2016), Vol. 198.
49. M. J. Bastiaans, "Transport Equations for the Wigner Distribution Function," Optica Acta: International Journal of Optics **26**, 1265–1272 (1979).
50. A. W. Lohmann, "Image rotation, Wigner rotation, and the fractional Fourier transform," J. Opt. Soc. Am. A **10**, 2181–2186 (1993).
51. E. Tseng, S. Colburn, J. Whitehead, L. Huang, S.-H. Baek, A. Majumdar, and F. Heide, "Neural nano-optics for high-quality thin lens imaging," Nat. Commun. **12**, 6493 (2021).
52. J. R. Shewell and E. Wolf, "Inverse diffraction and a new reciprocity theorem," J. Opt. Soc. Am. **58**, 1596–1603 (1968).
53. B. M. Hennelly and J. T. Sheridan, "Generalizing, optimizing, and inventing numerical algorithms for the fractional Fourier, Fresnel, and linear canonical transforms," J. Opt. Soc. Am. A **22**, 917–927 (2005).

54. X. Zhou, Q. Song, X. Yang, and W. Cai, "Generating phase-only diffractive optical elements using adaptive constraints in the Fourier domain," Opt. Commun. **535**, 129360 (2023).